\documentclass[aps, prd, twocolumn, showpacs, floatfix, letterpaper,
  nofootinbib, superscriptaddress,longbibliography]{revtex4-2}

\usepackage{graphicx}
\usepackage{epsfig}
\usepackage{bm}
\usepackage{pgf}
\usepackage{color}
\usepackage[T1]{fontenc}
\usepackage[latin1]{inputenc}
\usepackage{graphicx}
\usepackage[english]{babel}
\usepackage{amsmath}
\usepackage{amssymb}
\usepackage{amsfonts}
\usepackage{makecell}
\usepackage{hyperref}
\usepackage[draft,deletedmarkup=xout]{changes}
\usepackage{mathtools}
\usepackage{soul}
\usepackage{ulem}
\usepackage{relsize}
\usepackage{bbm}

\definecolor{green}{rgb}{0.19,0.64,0.54}
\definecolor{blue}{rgb}{0,0,1}
\definecolor{reddish}{rgb}{0.65, 0.2,0.2}
\definecolor{darkgreen}{rgb}{0.2,0.7,0.3}
\definecolor{darkblue}{rgb}{0.3,0.40,0.48}
\definecolor{gray}{rgb}{.8,.8,.8}

\newcommand{\scalar}[2]{\langle\kern.3ex #1
  \kern.3ex|\kern.3ex#2\kern.3ex\rangle}

\hypersetup{,
  pdftitle={Quantumness quantum cosmo},
  pdfauthor={H. Bergeron, P Malkiewicz and P Peter},
  colorlinks=true, 
  linkcolor=darkblue, 
  citecolor=darkgreen, 
  filecolor=reddish, 
  urlcolor=blue, 
  linktocpage=true
}

\newcommand{\dd}{\mathrm{d}}

\newcommand{\Rez}{\Re\mathrm{e}}

\newcommand{\eqdef}{\stackrel{\mbox{\tiny def}}{=}}  
\newcommand{\ket}[1]{|\kern.3ex#1\kern.3ex\rangle}
\newcommand{\ketn}[1]{\kern.3ex#1\kern.3ex\rangle}
\newcommand{\bra}[1]{\langle\kern.3ex #1 \kern.3ex|}
\newcommand{\norm}[1]{\|\kern.3ex#1\kern.3ex \|}

\begin{document}

\title{Non-Gaussianities as a Signature of Quantumness of Quantum
  Cosmology}

\author{Herv\'e Bergeron}

\email{herve.bergeron@universite-paris-saclay.fr}

\affiliation{Institut des Sciences Mol\'{e}culaires d'Orsay (ISMO),
  UMR 8214 CNRS, Universit\'{e} Paris-Saclay, 91405 Orsay Cedex,
  France}

\author{Przemys{\l}aw Ma{\l}kiewicz}
\email{Przemyslaw.Malkiewicz@ncbj.gov.pl}

\affiliation{National Centre for Nuclear Research, Pasteura 7, 02-093
  Warszawa, Poland}

\author{Patrick Peter}
\email{peter@iap.fr}

\affiliation{${\cal G}\mathbb{R}\varepsilon\mathbb{C}{\cal O}$ --
  Institut d'Astrophysique de Paris, CNRS and Sorbonne Universit\'e,
  UMR 7095 98 bis boulevard Arago, 75014 Paris, France}

\begin{abstract}
We show that the consistent application of the rules of quantum
mechanics to cosmological systems inevitably results in the so-called
multiverse states in which neither the background spacetime nor the
inhomogeneous perturbation are in definite states. We study the
multiverse states as perturbations to the usually employed so-called
Born-Oppenheimer states that are products of a wave function of the
background and a wave function of the perturbation. The obtained
corrections involve integrals over \emph{virtual backgrounds} that
represent the effect of quantum background fluctuations on the
perturbation state. They resemble loop corrections in quantum field
theory. This approach demonstrates the inevitable existence of very
specific non-Gaussian features in primordial fluctuations. We express
the resulting non-Gaussian perturbation as a nonlinear function of the
Gaussian perturbation obtained within the Born-Oppenheimer
approximation, and compute its trispectrum, to show that the
multiverse scenario leads to testable and distinct signatures in
cosmological perturbations. Our approach applies both to inflationary
and alternative cosmologies.
\end{abstract}

\maketitle

\section{Introduction}

Virtually all quantum cosmological
models~\cite{Hartle:1983ai,Vilenkin:1983xq, Halliwell:1984eu} that are
discussed in the literature assume the quantum state of the universe
to be the tensor-product of wavefunctions for the background mode and
the perturbation modes (see, for instance, Refs.~\cite{Pinho:2006ym,
  Peter_2006, Peter:2008qz, Lehners:2008vx,
  Ma_kiewicz_2021,Pinto-Neto:2013toa,PhysRevD.79.064030, Gomar_2014,
  KAMENSHCHIK2013518, PhysRevD.103.066005, PhysRevD.87.104008}). We
shall refer to this assumption as the Born-Oppenheimer (BO)
factorization. It is important to note that this assumption is highly
restrictive, as such states, called in what follows {\it
  Born-Oppenheimer (BO) states}, occupy a set of measure-zero in the
space of all possible quantum gravity states, even though some
justification can be provided from a decoherence point of
view~\cite{Kiefer:2007ria}. Despite the fact that such quantum states
can make some observationally consistent predictions for primordial
perturbations, the respective quantum cosmological models may seem
overly simplistic and, to some extent, questionable. Such a viewpoint
is particularly legitimate in case of the models that address the
initial singularity problem.

Apart from computational simplicity, there is no inherent reason to
assume that the universe should adhere to the BO factorization. On the
contrary, had the quantum effects been really significant in the
primordial universe, it is natural to expect that the underlying state
would not respect this artificial factorization. Indeed, as our
previous study \cite{Bergeron:2024art} showed, a simple bouncing model
typically evolves an initial BO state into a more generic one. The
latter is an entangled state that cannot be expressed as a product
state of the background and the perturbation but rather as a sum,
possibly infinite, of such products. In other words, neither the
background geometry nor any perturbation mode emerges from the big
bang in a definite state. Therefore, the conventional splitting of
cosmological spacetime into the background and the perturbation, which
can be disputed at the classical level \cite{Ellis:1987zz}, turns out
to be dissolved into entangled states by quantum gravity. Despite that
entangled states arise from quantization of a single-universe model,
we call them the {\it multiverse states} as each of them carries the
potentiality of multiple cosmological scenarios. It is noteworthy that
a consistent application of the rules of quantum mechanics to general
relativity inevitably results in the emergence of this multiverse
framework.

The idea of the multiverse is very attractive in the view of how
finely-tuned our Universe appears \cite{Livio:2018zqr}. It posits that
our cosmos is just one among numerous other universes that are
governed by the same fundamental laws but admit different initial
conditions \cite{Tegmark:1997in}. Although, the multiverse idea is not
new, it has never been given, to the best of our knowledge, the
specific form described above. In the context of inflationary models,
the multiverse refers to multiple, disconnected, expanding regions,
characterized by different cosmological parameters and existing within
a single and well-defined spacetime, even if vast or infinite
\cite{Linde:2015edk}. In the context of the many-world interpretation
of quantum mechanics \cite{Everett:1957hd}, the multiverse is indeed
made of multiple branches of the wave function of the Universe, but
the branching is usually understood as a process of splitting the
worlds by means of many small variations in local degrees of freedom,
and the discussions thereof do not normally involve the perturbed
cosmological model studied herein. Our results are, moreover,
independent of any specific interpretation of, or approach to, quantum
mechanics.

Apart from the fact that the multiverse idea can help explain the
fine-tuning of the initial condition, the significance of our specific
multiverse model lies in the fact that it makes experimentally
testable predictions. This is contrary to the common perception of the
multiverse as an interesting but fundamentally untestable idea. We
prove that the different branches of the multiverse state can
interact, leaving a distinct imprint of this interaction in the
primordial structure of the Universe. Interestingly, one of the
predicted multiverse signatures, the non-Gaussianity of the primordial
density fluctuations, is produced via interaction of the perturbation
modes with the quantum background rather than higher-order
interactions among the perturbation modes themselves. This makes our
prediction distinct from and independent of any specific higher-order
extension to the linear cosmological perturbation theory.

With this work, we develop the multiverse framework initiated in our
previous work \cite{Bergeron:2024art} where a coarse-grained approach
to the multiverse dynamics was established. Presently, we focus on
developing a perturbative approach based on the expansion of a generic
state of the universe in terms of the BO states. With this new
approach, as will become clearer later, we circumvent any potential
stability issue, likely present in the complete multiverse
dynamics. In what follows we make no assumption on the details of the
underlying quantum model, which results in a broad framework that is
applicable both to inflationary and alternative (e.g., big-bounce)
models. The only restriction we assume is that the states must be
perturbatively close to the BO states.

This new approach confirms and extends our previous result on the
existence of non-Gaussianity in primordial fluctuations as a direct
consequence of the entangling dynamics. Actually, it enables to pin
down the shape and the amplitude of non-Gaussianity via a four-point
correlation function. Thus, our result lays the foundations for
experimental verification of the multiverse scenario.

The plan of this work is as follows. In Sec.~\ref{QUsec}, we discuss
the idea of a generic quantum state of the universe together with the
Born-Oppenheimer-like restriction imposed on it in the usual approach
to quantum cosmology. Sec.~\ref{Projsec} introduces a useful tool
which is a projection operator, and shows how it applies to the
discussions of both the Born-Oppenheimer and the multiverse
states. Sec.~\ref{PTsec} derives the main result of this work, which
is the formula \eqref{BBOfinal1} for the multiverse state from a
perturbation theory. Sec.~\ref{PNGsec} uses the result of the
preceding section for computation of the four-point correlation
function that is given a specific analytical form. We summarize our
work in Sec. \ref{Sumsec}.

\section{Quantum Universe}
\label{QUsec}

The general framework of quantum postulates splits physical world into
(a) the quantized system which is under study, and for which no
``objective" properties can be assumed to exist, and (b) an external
domain, or the ``observer", for which unique properties of the studied
system exist, and are statistically predictable with the probabilities
obtainable from the quantum formalism. However, these postulates
cannot be directly applied to our entire Universe that (a) is unique
by definition, and (b) does not possess any ``external domain", or
equivalently, (b') internally includes any potential observer.

In this work, instead of trying to solve this outstanding issue, we
take an approach allowing us to use the standard Hilbert space
formalism and make physical predictions in a consistent way. We
explain our way of applying quantum formalism below.

\subsection{Quantum Cosmological Background} 

We shall assume that only one time-dependent state of a cosmological
background, denoted by $\ket{\alpha_\textsc{b}}$, represents our
Universe. Furthermore, we assume that the state
$\ket{\alpha_\textsc{b}}$ satisfies the background Schr\"{o}dinger
equation at all times,
$$
i\partial_{\eta}\ket{\alpha_\textsc{b}}=\hat{H}^{(0)}\ket{\alpha_\textsc{b}},
$$
so that the quantum background evolution of our Universe corresponds
to a fixed, specific state-trajectory,\footnote{We proved in
Ref.~\cite{Bergeron:2023zzo} that coherent states
$\ket{\alpha_\textsc{b}}=\ket{q,p}$ exist that evolve parametrically,
i.e. $\ket{\alpha_\textsc{b}(\eta)}=\ket{q_\eta,p_\eta}$, under the
action of $\hat{H}^{(0)}$. They are very good candidates to specify
state-trajectories.}
$$
\eta\mapsto \ket{\alpha_\textsc{b}(\eta)},
$$
where $\eta$ is the conformal time as measured with respect to an
arbitrary origin.

When a specific state-trajectory is chosen to represent our Universe,
all other states, denoted as $\ket{\alpha}$, must be seen as ``virtual
states" corresponding to potential configurations that the system
could assume, but in fact it does not. A quantum jump from one
trajectory to another is prohibited because there is no possible
measurement performed by an apparatus outside our Universe, producing
a non-unitary evolution of $\ket{\alpha_\textsc{b}}$, and thereby
breaking the Schr\"{o}dinger equation. However, coupling the
perturbation modes to the background geometry generically causes the
quantum background to spread into all the other branches
$\ket{\alpha}$, with each branch in turn influencing the perturbation
modes in a distinctive manner. Although the virtual background states
remain physically inaccessible, their effect on the perturbation modes
is potentially observable and thus, physically relevant.

In the following, we assume that the set of states $\{\ket{\alpha}\}$
(including $\ket{\alpha_\textsc{b}}$) resolves the identity, i.e.,
$$
\int \dd \alpha
\ket{\alpha}\bra{\alpha} =\mathbbm{1},
$$
and forms a basis for the background Hilbert space
$\mathcal{H}_\textsc{b}$.

\subsection{Virtual Backgrounds}

The point of view taken in this work suggests that the only observable
background quantities are the expectation values of any quantum
operators in the state $\ket{\alpha_\textsc{b}}$, understood as
"classical" quantities, e.g., the scale factor. Note that this implies
that the usually postulated observables of ordinary quantum mechanics,
that is, the eigenvalues of real operators, are in general excluded
from the proposed formalism.

Instead, we can predict the expectations values
$\bra{\alpha_\textsc{b}(\eta)} \hat{\mathcal{O}}
\ket{\alpha_\textsc{b}(\eta)}$ of any observable $\hat{\mathcal{O}}$
as functions of time in the background state of interest
$\ket{\alpha_\textsc{b}(\eta)}$. The effect of virtual states appear
when, for example, we compute the expectation value of a compound
observable such as $\bra{\alpha_\textsc{b}} \hat{\mathcal{O}}^2
\ket{\alpha_\textsc{b}}$, which, thanks to the resolution of the
identity, is given by
\begin{align}
\bra{\alpha_\textsc{b}} \hat{\mathcal{O}}^2 \ket{\alpha_\textsc{b}} =
\int \dd \alpha \bra{\alpha_\textsc{b}} \hat{\mathcal{O}}
\ket{\alpha}\bra{\alpha} \hat{\mathcal{O}} \ket{\alpha_\textsc{b}}.
\end{align}
This actually differs from $\left( \bra{\alpha_\textsc{b}}
\hat{\mathcal{O}} \ket{\alpha_\textsc{b}} \right)^2$, and instead, one
often gets
\begin{align}
 \bra{\alpha_\textsc{b}} \hat{\mathcal{O}}^2 \ket{\alpha_\textsc{b}} =
 \lambda \left(\bra{\alpha_\textsc{b}} \hat{\mathcal{O}}
 \ket{\alpha_\textsc{b}} \right)^2,
\end{align}
\noindent 
where the renormalization factor $\lambda$ can be interpreted as the
effect of virtual states $\ket{\alpha}$. It is similar (but not
completely equivalent) to loop corrections in quantum field theory.

\subsection{Perturbing the background}

The classical Hamiltonian can be obtained from the ADM formalism by
truncating it at second order of perturbations and solving both the
background-level and the linearized constraints. This procedure
involves the choice of some internal time, which for the purpose of
this study has been conveniently set to coincide with conformal time
(see our previous work \cite{Bergeron:2024art} for more details). The
validity of the truncation rests on the assumption of smallness of the
perturbations to homogeneity. For a perturbed flat FLRW universe
filled with a perfect fluid, we have
\begin{equation}
H(q,p,v,\pi) = H^{(0)}(q,p) +
\sum^{\bm{k}_\text{max}}_{\bm{k}=\bm{k}_\text{min}}
H^{(2)}_{\bm{k}}(q,p,v_{\bm{k}},\pi_{\bm{k}}),
\label{globalclassH}
\end{equation}
where $(q,p)$ are canonical background variables and
$$
H^{(2)}_{\bm{k}} = |\pi_{\bm{k}}|^2 +\left[{\bm{k}}^2-
  V_\textsc{b}(q,p) \right] |v_{\bm{k}}|^2
$$
describes the dynamics of the are canonical perturbation variables
$(v_{\bm{k}},\pi_{\bm{k}})$ for the Fourier modes ${\bm{k}}$. Note
that, the perturbations being real, we have $|v_{\bm{k}}|^2 =
v_{\bm{k}} v^*_{\bm{k}} = v_{\bm{k}}v_{-\bm{k}}$; this relation will
still hold in the quantum context below. We also assume that our
perturbation model \eqref{globalclassH} represents an effective field
theory and is thus only valid for a limited range of wave vectors
$\bm{k} \in [\bm{k}_\text{min},\bm{k}_\text{max}]$\footnote{Actually,
in standard cosmology with slightly red power spectrum ($\mathcal{P}_k
\propto k^{n_\textsc{s}-1}$ with $n_\textsc{s} <1$), setting
$|\bm{k}|_\text{max}<\infty$ is not necessary, whereas
$|\bm{k}|_\text{min}>0$ reflects the finite size of the studied
spacetime region.}.

The truncated Hamiltonian $H(q,p, v, \pi)$ is physically correct,
i.e., consistent with the initial assumption, only if the Hamilton
equations for the background deduced from $H(q,p, v, \pi)$ are almost
the same as those obtained from $H^{(0)}$. Or put differently, the
inevitable (for $V_\textsc{b} \neq 0$) dynamical backreaction of the
perturbation variables $(v, \pi)$ on the background $(q,p)$, induced
by $H$, must always remain very weak. But in fact we see that the
backreaction is induced by the change
\begin{equation}
H^{(0)}(q,p) \mapsto H^{(0)}(q,p)- V_\textsc{b}(q,p)
\sum^{\bm{k}_\text{max}}_{\bm{k}=\bm{k}_\text{min}} |v_{\bm{k}}|^2.
\end{equation}
Therefore, the backreaction is negligible only if
$V_\textsc{b}(q,p)\sum_{\bm{k}} |v_{\bm{k}}|^2 \ll 1$, which is
consistent with our assumption that the theory is meaningful only in a
limited range of $\bm{k}$. Far from a singularity ($q=0$), this
constraint can certainly be satisfied and dynamically preserved, i.e.,
made valid uniformly in time. However, on the approach to the
singularity, it cannot be ensured that the backreaction remains
weak. Indeed, sufficiently close to the singularity, even if the
linear Hamilton equations for the perturbations variables $(v, \pi)$
are assumed to remain valid, the background variables $(q,p)$ may no
longer satisfy the zeroth-order equations of motion.

One could conclude that the perturbation framework is simply not valid
close to the singularity, and that the calculations cannot be trusted
in this regime. However, there exists a mathematically consistent way
to keep the perturbation framework by explicitly removing the
backreaction The idea is to solve the Hamilton equations in two steps:

(a) we first specify a background trajectory $(q_\eta, p_\eta)$ only
by using the background-level Hamiltonian $H^{(0)}(q,p)$,

(b) then we find the evolution of the perturbation variables
$v_k(\eta)$ and $\pi_k(\eta)$ by using the time-dependent Hamiltonian
$H^{(2)}(q_\eta,p_\eta,v_k,\pi_k)$. This is in fact the usual approach
to perturbation theory at the classical level.

The quantum counterpart of the classical system involves two partial
Hilbert spaces: $\mathcal{H}_\textsc{b}$ for the background and
$\mathcal{H}_\textsc{p}$ for the perturbations. The total Hilbert
space is therefore the tensor product $\mathcal{H} =
\mathcal{H}_\textsc{b} \otimes \mathcal{H}_\textsc{p}$. We have a
quantum Hamiltonian $\hat{H}^{(0)}$ acting only on
$\mathcal{H}_\textsc{b}$ for the background and a quantum perturbation
Hamiltonian $\hat{H}^{(2)}$ acting on $ \mathcal{H}_\textsc{b} \otimes
\mathcal{H}_\textsc{p}$. In order to avoid any kind of quantum
backreaction from the perturbation modes on the quantum background, we
need to follow a two-step procedure similar to the classical one. This
leads us to {\it the Born-Oppenheimer factorization}, as we explain
below.

\subsection{Born-Oppenheimer factorization}

Following the previous analysis of the quantum background, we first
specify a trajectory-state $\ket{\alpha_\textsc{b}(\eta)}$ that
follows the Schr\"{o}dinger equation. This corresponds to the step (a)
of the classical procedure. The quantum counterpart of the step (b) is
given by the solution of the following Schr\"{o}dinger equation for
the perturbation modes
\begin{equation}
i \partial_\eta \ket{\psi_\textsc{p}(\eta)} = \tilde{H}^{(2)}
\ket{\psi_\textsc{p}(\eta)}
\end{equation}
with
\begin{equation}
\tilde{H}^{(2)}=\bra{\alpha_\textsc{b}(\eta)}
\hat{H}^{(2)}\ket{\alpha_\textsc{b}(\eta)},
\end{equation}
where $\tilde{H}^{(2)}$ represents the operator $\hat{H}^{(2)}$
averaged on the background state
$\ket{\alpha_\textsc{b}(\eta)}$. Therefore, even though it may look as
a scalar product, the quantity $\tilde{H}^{(2)}$ remains an operator,
but one that acts only on the perturbation space
$\mathcal{H}_\textsc{p}$. This procedure involves a factorized global
state $\ket{\alpha_\textsc{b}(\eta)} \otimes
\ket{\psi_\textsc{p}(\eta)}$ which we refer to as {\it the
  Born-Oppenheimer state}.

\subsection{Beyond Born-Oppenheimer factorization}

Quantum formalism introduces physical effects that have no classical
counterparts such as quantum entanglement. It turns out that it is
possible to go beyond the Born-Oppenheimer factorization and find the
effects of entanglement while staying within a quantum counterpart of
the previously described two-step procedure.

The idea is that the virtual background states, thanks to the {\it
  background dynamical loops}, are able to produce new effects that
have no equivalent in the classical domain, but are neglected by the
Born-Oppenheimer factorization. In other words, the background quantum
fluctuations are able to modify the Born-Oppenheimer states. The
corresponding corrections are similar to the loop corrections in
quantum field theory.

Let us define the complete (time-independent) Hamiltonian as
\begin{equation}
\hat{H} = \hat{H}^{(0)} + \hat{H}^{(2)},
\label{hamil0}
\end{equation}
and let us consider a general state $\ket{\Phi (\eta)} \in
\mathcal{H}$, which satisfies the full Schr\"{o}dinger equation:
\begin{equation}
i \partial_\eta \ket{\Phi (\eta)} = \hat{H} \, \ket{\Phi (\eta)}.
\label{schro0}
\end{equation}
The state $\ket{\Phi (\eta)}$ can be interpreted as a \emph{multiverse
state}, as it potentially involves a superposition of different
Born-Oppenheimer states. Indeed, even if the evolution starts from a
factorized state at time $\eta_0$, i.e., $\ket{\Phi (\eta_0)} =
\ket{\alpha_\textsc{b}(\eta_0)} \otimes
\ket{\psi_\textsc{p}(\eta_0)}$, at later times this state develops
into a multiverse state $\ket{\Phi(\eta)}$. This is so because the
action of the Hamiltonian $\hat{H}^{(2)}$ involves mixing between
different background states (i.e., $\hat{H}^{(2)}$ is in general
non-diagonal in states $\ket{\alpha}$). Furthermore, the multiverse
state $\ket{\Phi(\eta)}$ includes the quantum backreaction from the
perturbation on the background, which is inevitable and may grow
uncontrollably, and the assumption of its weakness could break
down. This is similar to the classical situation where we directly
apply the complete classical Hamiltonian $H$ of
Eq. \eqref{globalclassH}.

Starting from the Schr\"{o}dinger equation \eqref{schro0}, at first
sight it seems impossible to respect the constraint of the weak
backreaction, and the Born-Oppenheimer factorization seems to be the
only way to implement this constraint at the quantum level.

But this conclusion is too rash, and in fact a generic multiverse
state should be considered. We find that imposing correctly the
weak-backreaction constraint at the quantum level involves three
steps:

(a) solve the complete dynamical equations and determine the entangled
multiverse state $\ket{\Phi (\eta)}$ for an initial factorized state
$\ket{\Phi (0)}=\ket{\alpha_\textsc{b}(0)} \otimes
\ket{\psi_\textsc{p}(0)}$,

(b) determine the background state-trajectory
$\ket{\alpha_\textsc{b}(\eta)}$ that evolves according to the
background Hamiltonian $\hat{H}^{(0)}$,

(c) obtain the perturbation state proper to the chosen background
$\ket{\alpha_\textsc{b}(\eta)}$ by calculating the partial scalar
product
\begin{equation}
\ket{\psi_\textsc{p}(\eta)} \eqdef \frac{1}{Z_\eta}
\scalar{\alpha_\textsc{b}(\eta)}{\Phi(\eta)}\,\rangle_{\mathcal{H}_\textsc{p}},
\label{perturb1}
\end{equation}
where $Z_\eta$ is introduced for the normalization, and we added an
extra $\,\rangle_{\mathcal{H}_\textsc{p}}$ to emphasize the result,
although looking like a scalar product, is in fact a vector living in
$\mathcal{H}_\textsc{p}$. In terms of integrals, we have
\begin{equation}
\psi_\textsc{p}(v_k,\eta) \eqdef \frac{1}{Z_\eta} \int \dd \alpha \,
\scalar{\alpha_\textsc{b}(\eta)}{\alpha} \Phi(\alpha,v_k, \eta).
\end{equation}
We note that this formula does not introduce any new effects in the
case of the Born-Oppenheimer state: if $\ket{\Phi(\eta)} =
\ket{\Phi^{(\textsc{bo})}(\eta)} = \ket{\alpha_\textsc{b}(\eta)}
\otimes \ket{\psi_\textsc{p}^{(\textsc{bo})}(\eta)}$, then
$\scalar{\alpha_\textsc{b}(\eta)}{\Phi^{(\textsc{bo})}(\eta)}
\,\rangle_{\mathcal{H}_\textsc{p}} =
\ket{\psi_\textsc{p}^{(\textsc{bo})}(\eta)}$. But since the
Hamiltonian $\hat{H}$ of Eq. \eqref{hamil0} spreads the
Born-Oppenheimer state over the multiverse, the above formula
generically yields a new definition of the perturbation state.

To summarize, we note that defining the perturbation state as
$\ket{\psi_\textsc{p}(\eta)}$ in Eq. \eqref{perturb1}, satisfying the
initial Born-Oppenheimer factorization $\ket{\Phi
  (\eta_0)}=\ket{\alpha_\textsc{b}(\eta_0)} \otimes
\ket{\psi_\textsc{p}(\eta_0)}$, allows us to:

(a) respect the constraint of negligible backreaction by setting a
specific background solution evolving according to the
background-level Hamiltonian $\hat{H}^{(0)}$. We first include the
undesired backreaction effects using the full Hamiltonian $\hat{H}$,
and then we remove them by projection and renormalization, keeping
only the effects of $\hat{H}$ on the perturbation state,

(b) take into account the dynamical effect of all the virtual
backgrounds on the perturbation state, which is neglected by the
Born-Oppenheimer factorization,

(c) obtain finally a factorized state $\ket{\alpha_\textsc{b}(\eta)}
\otimes \ket{\psi_\textsc{p}(\eta)}$ as the {\it observed state} that
is free of the multiverse ambiguity.

As the dominant part of $\ket{\psi_\textsc{p}(\eta)}$ is given by the
Born-Oppenheimer perturbation state, the multiverse corrections can be
calculated perturbatively and, as we show below, they generically lead
to non-Gaussian perturbations.

\section{Technical preliminary}
\label{Projsec}

\subsection{Projection onto a background state}

Our goal is to develop a framework for calculating the multiverse
dynamics that violates the Born-Oppenheimer factorization. In order to
obtain a concise formulation we introduce technical tools allowing us
to shorten the expressions while avoiding ambiguities.

For each normalized state $\ket{\alpha_\textsc{b}} \in
\mathcal{H}_\textsc{b}$ of the background we associate a projector
$\hat{\Pi}_{\alpha_\textsc{b}}: \mathcal{H} \mapsto
\mathcal{H}_\textsc{p}$ acting on the full Hilbert space $\mathcal{H}=
\mathcal{H}_\textsc{b} \otimes \mathcal{H}_\textsc{p}$ towards the
perturbation Hilbert space $\mathcal{H}_\textsc{p}$, defined as
\begin{equation}
\hat{\Pi}_{\alpha_\textsc{b}} \eqdef \sum_{n_\textsc{p}}
\ket{n_\textsc{p}} \, \left( \bra{\alpha_\textsc{b}} \otimes
\bra{n_\textsc{p}} \right),
\label{proj0}
\end{equation}
where $\{ \ket{n_\textsc{p}} \}$ is any orthonormal basis of
$\mathcal{H}_\textsc{p}$. It is easy to prove that
$\hat{\Pi}_{\alpha_\textsc{b}}$ is not dependent on the choice of the
basis $\{ \ket{n_\textsc{p}} \}$ of $\mathcal{H}_\textsc{p}$, and
depends only on $\ket{\alpha_\textsc{b}}$. This projector allows to
define a (non-normalized) perturbation state $\ket{\psi_\textsc{p}} =
\hat{\Pi}_{\alpha_\textsc{b}} \ket{\Phi}$ that corresponds to a given
background state $\ket{\alpha_\textsc{b}}$ in the entangled multiverse
state $\ket{\Phi} \in \mathcal{H}_\textsc{b} \otimes
\mathcal{H}_\textsc{p}$.

We can also define its adjoint $\hat{\Pi}_{\alpha_\textsc{b}}^\dagger:
\mathcal{H}_\textsc{p} \mapsto \mathcal{H} $ as
\begin{equation}
\hat{\Pi}_{\alpha_\textsc{b}}^\dagger = \sum_{n_\textsc{p}}
(\ket{\alpha_\textsc{b}} \otimes \ket{n_\textsc{p}}) \,
\bra{n_\textsc{p}},
\label{proj1}
\end{equation}
and the operators $\hat{\Pi}_{\alpha_\textsc{b}}$ and
$\hat{\Pi}_{\alpha_\textsc{b}}^\dagger $ verify the relations
\begin{align}
\begin{split}
\hat{\Pi}_{\alpha_\textsc{b}}^\dagger \hat{\Pi}_{\alpha_\textsc{b}} &=
(\ket{\alpha_\textsc{b}} \, \bra{\alpha_\textsc{b}}) \otimes
\mathbb{I}_{\mathcal{H}_\textsc{p}},\\ \hat{\Pi}_{\alpha_\textsc{b}}
\hat{\Pi}_{\alpha_\textsc{b}}^\dagger &=
\mathbb{I}_{\mathcal{H}_\textsc{p}}.
\end{split}
\label{proj2}
\end{align}
More generally, for any operators $\hat{W}_\textsc{b}$ and
$\hat{X}_\textsc{p}$, we have
\begin{align}
\hat{\Pi}_{\alpha_\textsc{b}'}^\dagger \hat{X}_\textsc{p}
\hat{\Pi}_{\alpha_\textsc{b}} &=
(\ket{\alpha_\textsc{b}'}\bra{\alpha_\textsc{b}}) \otimes
\hat{X}_\textsc{p},\label{projmix1}\\ \hat{\Pi}_{\alpha_\textsc{b}}
\left( \hat{W}_\textsc{b} \otimes \hat{X}_\textsc{p} \right)
\hat{\Pi}_{\alpha_\textsc{b}'}^\dagger &= \bra{ \alpha_\textsc{b}}
\hat{W}_\textsc{b} \ket{\alpha_\textsc{b}'} \, \hat{X}_\textsc{p},
\label{projmix}
\end{align}
in which $\hat{W}_\textsc{b}$ and $\hat{X}_\textsc{p}$ respectively
act on $\mathcal{H}_\textsc{b}$ and $\mathcal{H}_\textsc{p}$.

\subsection{Projected multiverse state}

It is easy to note that, given an orthonormal basis of background
states $\{ \ket{n_\textsc{b}} \}$ of $\mathcal{H}_\textsc{b}$,
Eq. \eqref{proj2} implies
\begin{equation}
\sum_{n_\textsc{b}} \hat{\Pi}_{n_\textsc{b}}^\dagger
\hat{\Pi}_{n_\textsc{b}} = \left(\sum_{n_\textsc{b}}
\ket{n_\textsc{b}} \, \bra{n_\textsc{b}} \right) \otimes
\mathbb{I}_{\mathcal{H}_\textsc{p}}=
\mathbb{I}_{\mathcal{H}_\textsc{b}} \otimes
\mathbb{I}_{\mathcal{H}_\textsc{p}} = \mathbb{I}_{\mathcal{H}}.
\label{proj3}
\end{equation}
Let us assume that the cosmological system at a given time is in a
normalized \emph{multiverse state} $\ket{\Phi} \in \mathcal{H}$. Then,
according to the Born rule and Eq. \eqref{proj2},
$P(\alpha_\textsc{b}) = \bra{\Phi}
\hat{\Pi}_{\alpha_\textsc{b}}^\dagger \hat{\Pi}_{\alpha_\textsc{b}}
\ket{\Phi}=\bra{\Phi} (\ket{\alpha_\textsc{b}} \,
\bra{\alpha_\textsc{b}}) \otimes \mathbb{I}_{\mathcal{H}_\textsc{p}}
\ket{\Phi}$ is the probability to find the system in the normalized
background state $\alpha_\textsc{b}$. Given an orthonormal basis $\{
\ket{n_\textsc{b}} \}$ of $\mathcal{H}_\textsc{b}$, we obtain that the
universe is found in some background state with certainty (see Eq.
\eqref{proj3}),
\begin{equation}
\sum_{n_\textsc{b}} P(n_\textsc{b}) = \bra{\Phi}\Phi \rangle=1.
\end{equation}
Note that above we have used the standard language of probabilities,
in spite of the fact that it may be done only formally as the standard
interpretation of quantum formalism is not valid in the present case.
The concept of measurement is, in practice, not applicable to the
quantum theory of cosmological background, which involves the
multiverse states. Nevertheless, these {\it formal probabilities},
which cannot be verified by any measurement, can be computed.

If the cosmological system is found in the background state
$\ket{\alpha_\textsc{b}}$, then, from the postulate of the collapse of
the wave-function, the normalized perturbation state becomes
$$
\ket{\psi_\textsc{p}} = \frac{1}{\sqrt{P(\alpha_\textsc{b})}}
\hat{\Pi}_{\alpha_\textsc{b}} \, \ket{\Phi},
$$
and the ensuing full normalized state ends up being factorized, i.e.,
$$
\ket{\Phi'} = \ket{\alpha_\textsc{b}} \otimes \ket{\psi_\textsc{p}} =
\frac{1}{\sqrt{P(\alpha_\textsc{b})}}
~\hat{\Pi}_{\alpha_\textsc{b}}^\dagger\hat{\Pi}_{\alpha_\textsc{b}} \,
\ket{\Phi}.
$$
The expression for $\ket{\psi_\textsc{p}}$ provides the most general
perturbation state that includes features that are neglected in the
Born-Oppenheimer approach. It is not to be understood as the outcome
of a measurement, which would be meaningless in the case of the entire
Universe. Instead, it makes definite predictions concerning the
primordial structure {\it provided that} the Universe is in a given
background state. It is therefore assumed that there exists a
preferred background state that corresponds to the observable Universe
and exhibits a specific dynamical behavior governed by the background
Hamiltonian.

In what follows, we shall adopt a compact notation for the background
average of $\hat{H}^{(2)}$ between two background states, $\alpha$ and
$\beta$,
\begin{equation}
\tilde{H}^{(2)}_{\alpha,\beta} \eqdef \hat{\Pi}_{\alpha} \hat{H}^{(2)}
\hat{\Pi}^\dagger_{\beta} \equiv \bra{\alpha} \hat{H}^{(2)}
\ket{\beta},
\label{AvgH2}
\end{equation}
and we will further simplify the notation for the diagonal elements
(i.e., with $\alpha=\beta$) by setting $\tilde{H}^{(2)}_{\alpha}
\eqdef \tilde{H}^{(2)}_{\alpha, \alpha}$. As before, one should remark
that even though the left hand side of Eq.~\eqref{AvgH2} appears like
a scalar product, it is only partial, so the remaining quantity is
indeed an operator, acting on the perturbation Hilbert space
$\mathcal{H}_\textsc{p}$ only.

\subsection{Reformulation of the Born-Oppenheimer approach}

The background trajectory-state $\ket{\alpha_\textsc{b}(\eta)}$
satisfies the zeroth-order Schr\"{o}dinger equation
\begin{equation}
i \partial_\eta \ket{\alpha_\textsc{b}(\eta)} = \hat{H}^{(0)}
\ket{\alpha_\textsc{b}(\eta)},
\end{equation}
which has the solution
\begin{equation}
\ket{\alpha_\textsc{b}(\eta)} = e^{-i \hat{H}^{(0)} (\eta-\eta_0)}
\ket{\alpha_\textsc{b}(\eta_0)}.
\label{evol0}
\end{equation}
The Born-Oppenheimer perturbation state
$\ket{\psi_\textsc{p}^{(\textsc{bo})}(\eta)}$ satisfies the
second-order Schr\"{o}dinger equation
\begin{equation}
i \partial_\eta \ket{\psi_\textsc{p}^{(\textsc{bo})}(\eta)} =
\tilde{H}^{(2)}_{\alpha_\textsc{b}(\eta)}
\ket{\psi_\textsc{p}^{(\textsc{bo})}(\eta)},
\label{evol2}
\end{equation}
whose solution reads
\begin{equation}
\ket{\psi_\textsc{p}^{(\textsc{bo})}(\eta)} = T \left\{ \exp\left[ -i
  \int_{\eta_0}^\eta \tilde{H}^{(2)} _{\alpha_\textsc{b}(\tau)} \, \dd
  \tau \right]\right\} \ket{\psi_\textsc{p}^\textsc{(bo)}(\eta_0)},
\end{equation}
using the Dyson time ordering operator $T[.]$. The total BO state
$\ket{\Phi^{(\textsc{bo})}(\eta)}$ at any time $\eta$ reads
\begin{equation}
\ket{\Phi^{(\textsc{bo})}(\eta)} = \ket{\alpha_\textsc{b}(\eta)}
\otimes \ket{\psi_\textsc{p}^{(\textsc{bo})}(\eta)}
\end{equation}
Given the unitary evolution of the background
$\ket{\alpha_\textsc{b}(\tau)}$, we have
\begin{align}\begin{split}
\tilde{H}^{(2)}_{\alpha_\textsc{b}(\tau)}
&=\hat{\Pi}_{\alpha_\textsc{b}(0)} e^{i \hat{H}^{(0)} \tau}
\hat{H}^{(2)} e^{-i \hat{H}^{(0)} \tau}
\hat{\Pi}^\dagger_{\alpha_\textsc{b}(0)}\\ &=
\hat{\Pi}_{\alpha_\textsc{b}(0)} \hat{H}^{(2)}_\textsc{i}(\tau)
\hat{\Pi}^\dagger_{\alpha_\textsc{b}(0)},\end{split}
\end{align}
where $\hat{H}^{(2)}_\textsc{i}(\eta) = e^{i \hat{H}^{(0)} \eta}
\hat{H}^{(2)} e^{-i \hat{H}^{(0)} \eta}$ is the perturbation
Hamiltonian in the interaction picture and we consider an evolution
from the arbitrary reference point at $\tau=0$.  Thus, the interaction
picture underlies the BO approach.

The BO state $\ket{\Phi^{(\textsc{bo})}(\eta)}$ can be written as
\begin{equation}
\label{phiBOstate}
\ket{\Phi^{(\textsc{bo})}(\eta)} = \hat{U}^{(\textsc{bo})}(\eta,
\eta_0) \ket{\Phi^{(\textsc{bo})}(\eta_0)},
\end{equation}
where, given Eqs.~\eqref{evol0} and \eqref{evol2},
\begin{equation}
\hat{U}^{(\textsc{bo})}(\eta, \eta_0) = T \left( \exp\left\{ -i
\int_{\eta_0}^\eta \left[ \hat{H}^{(0)} +
  \tilde{H}^{(2)}_{\alpha_\textsc{b}(\tau)} \right] \, \dd \tau
\right\} \right),
    \label{unitBO}
\end{equation}
which is valid because because $\hat{H}^{(0)}$ and
$\tilde{H}^{(2)}_{\alpha_\textsc{b}(\tau)}$ commute due to their
independent action on the background Hilbert space by $\hat{H}^{(0)}$
and on the perturbation Hilbert space by
$\tilde{H}^{(2)}_{\alpha_\textsc{b}(\tau)}$.  The unitary evolution of
$\ket{\Phi^{(\textsc{bo})}(\eta)}$, generated by
$\hat{U}^{(\textsc{bo})}(\eta, \eta_0)$, is in fact non-linear as
$\hat{U}^{(\textsc{bo})}(\eta, \eta_0) $ depends on the background
state $\ket{\alpha_\textsc{p}(\eta)}$.\\ The state
$\ket{\Phi^{(\textsc{bo})}(\eta)}$ is a specific solution of the
non-linear Schr\"{o}dinger equation
\begin{equation}
i \partial_\eta \ket{\Phi^{(\textsc{bo})}(\eta)} =\left[\hat{H}^{(0)}+
  \tilde{H}^{(2)}_{\alpha_\textsc{b}(\eta)} \right]
\ket{\Phi^{(\textsc{bo})}(\eta)},
\end{equation}
which is non-linear by virtue of the background dependence of
$\hat{H}^{(2)}_{\alpha_\textsc{b}(\eta)}$.

\section{Multiverse dynamics}
\label{PTsec}

\subsection{Exact dynamics}

The multiverse state $\ket{\Phi(\eta)}$ satisfies the Schr\"{o}dinger
equation
\begin{equation}
i \partial_\eta \ket{\Phi(\eta)} = \left[ \hat{H}^{(0)}+\hat{H}^{(2)}
  \right] \ket{\Phi(\eta)},
\end{equation}
which has the solution
\begin{align}
\ket{\Phi(\eta)} = \hat{U}_\text{tot}(\eta,\eta_0) \ket{\Phi(\eta_0)},
\end{align}
where the evolution operator $ \hat{U}_\text{tot}$ reads
\begin{align}
\hat{U}_\text{tot}(\eta,\eta_0)= e^{-i \left[
    \hat{H}^{(0)}+\hat{H}^{(2)} \right] (\eta-\eta_0)},
\end{align}
since neither $\hat{H}^{(0)}$ nor $\hat{H}^{(2)}$ explicitly depend on
time.  The usual transformation to the interaction picture with a new
evolution operator $\hat{S}$ is given by
\begin{equation}
\hat{U}_\text{tot}(\eta,\eta_0) = e^{-i \hat{H}^{(0)} \eta}
\hat{S}(\eta,\eta_0) \, e^{i \hat{H}^{(0)} \eta_0}
\label{unitglob}
\end{equation}
where
$$
\hat{S}(\eta,\eta_0) = T \left\{ \exp \left[ -i \int_{\eta_0}^\eta
  \hat{H}^{(2)}_\textsc{i}(\tau) \dd \tau \right] \right\},
$$
and, as before, $\hat{H}^{(2)}_\textsc{i}(\tau) = e^{i \hat{H}^{(0)}
  \tau} \hat{H}^{(2)} e^{-i \hat{H}^{(0)} \tau}$.

\subsection{Connection between exact and BO dynamics}

In what follows, we assume the multiverse state $\ket{\Phi(\eta_0)}$ n
the remote past (i.e., for $\eta_0 \to -\infty$) to be factorized as
$\ket{\Phi(\eta_0)} = \ket{\alpha_\textsc{b}(\eta_0)} \otimes
\ket{\psi_\textsc{p}(\eta_0)}$. For inflationary models, this would
imply the background $\ket{\alpha_\textsc{b}(\eta_0)}$ to represent a
small and expanding universe, whereas for alternative models such as
bouncing cosmologies, the background would be large and contracting,
with all relevant modes inside the Hubble radius,
i.e. $k>|\mathcal{H}|$, in both cases.

Making use of the Born-Oppenheimer state
$\ket{\Phi^{(\textsc{bo})}(\eta)}$ given by Eq. \eqref{phiBOstate},
the initial state $\ket{\Phi(\eta_0)} =
\ket{\alpha_\textsc{b}(\eta_0)} \otimes \ket{\psi_\textsc{p}(\eta_0)}$
can be written as
\begin{equation}
\ket{\Phi(\eta_0)} = \left[ \hat{U}^{(\textsc{bo})}(\eta, \eta_0)
  \right]^\dagger \ket{\Phi^{(\textsc{bo})}(\eta)}.
\end{equation}
Therefore, the multiverse state $\ket{\Phi(\eta)} $ can be obtained
from the Born-Oppenheimer state,
\begin{equation}
\ket{\Phi(\eta)} =\hat{V}(\eta,\eta_0)
\ket{\Phi^{(\textsc{bo})}(\eta)},
\label{PhiEvolV}
\end{equation}
where
\begin{equation}
\hat{V}(\eta,\eta_0) =\hat{U}_\text{tot}(\eta,\eta_0) \left[
  \hat{U}^{(\textsc{bo})}(\eta, \eta_0) \right]^\dagger.
\label{Vdef}
\end{equation}
Note that, by construction, $\hat{V}(\eta_0,\eta_0) = \mathbbm{1}$ and
$\ket{\Phi(\eta_0)}= \ket{\Phi^{(\textsc{bo})}(\eta_0)}=
\ket{\alpha_\textsc{b}(\eta_0)} \otimes
\ket{\psi_\textsc{p}(\eta_0)}$.

By using Eqs \eqref{unitBO} and \eqref{unitglob}, we evaluate
$\hat{V}(\eta,\eta_0)$ as
\begin{widetext}
\begin{equation}
\hat{V}(\eta,\eta_0) = e^{-i \hat{H}^{(0)} \eta} \, T \left[e^{-i
    \int_{\eta_0}^\eta \hat{H}^{(2)}_\textsc{i}(\tau) \dd \tau}
  \right] e^{i \hat{H}^{(0)} \eta_0} e^{i \hat{H}^{(0)} (\eta-\eta_0)}
\, T \left[ e^{i \int_{\eta_0}^\eta
    \tilde{H}^{(2)}_{\alpha_\textsc{b}(\tau)} \, \dd \tau} \right],
\end{equation}
or, using the fact that $\hat{H}^{(0)}$ commutes with
$\tilde{H}^{(2)}_{\alpha_\textsc{b}(\tau)}$,
\begin{equation}
\hat{V}(\eta,\eta_0) = e^{-i \hat{H}^{(0)} \eta} \, T \left[e^{-i
    \int_{\eta_0}^\eta \hat{H}^{(2)}_\textsc{i}(\tau) \dd \tau}
  \right] T \left[ e^{i \int_{\eta_0}^\eta
    \tilde{H}^{(2)}_{\alpha_\textsc{b}(\tau)} \, \dd \tau} \right]
e^{i \hat{H}^{(0)} \eta}.
\label{Vfinal}
\end{equation}
\end{widetext}
We will next assume that the multiverse state always remains close to
the Born-Oppenheimer state, and therefore the action of
$\hat{V}(\eta,\eta_0)$ can be treated perturbatively.

\subsection{Perturbative expansion}

The perturbative expansion is valid only if the Born-Oppenheimer state
is a good approximation to the exact solution uniformly in
time. Equivalently, the non-diagonal terms $\tilde{H}^{(2)}_{\alpha
  (\tau), \beta(\tau)}$ in the Hamiltonian \eqref{Vfinal} have to
remain much smaller than the diagonal ones
$\tilde{H}^{(2)}_{\alpha(\tau)}$. This condition was indeed satisfied
by the example considered in our previous work
\cite{Bergeron:2024art}. If this condition is not fulfilled, then the
perturbative expansion, implicitly assuming the BO state as a
zeroth-order solution, becomes irrelevant. Note, however, that even if
the perturbative expansion fails, the exact formula \eqref{Vfinal} for
$\hat{V}(\eta, \eta_0)$ remains valid. Using the language of
multiverse, in the non-perturbative strongly mixing case, the Universe
becomes delocalized over all branches (i.e., background states with
exact perturbations). This implies that the formal probability to find
the Universe in any specific background state becomes nonvanishing but
small.

The development up to the second order of $\hat{V}(\eta,\eta_0)$
yields four terms, namely
\begin{align}
\hat{V}(\eta,\eta_0)\simeq e^{-i \hat{H}^{(0)} \eta} \left( 
\mathbb{I} + \hat{C}_1 + \hat{C}'_2 + \hat{C}''_2 \right)
e^{i \hat{H}^{(0)} \eta},
\label{perturV0}
\end{align}
where
\begin{widetext}
\begin{subequations}
\begin{align}
\hat{C}_1 &= -i \int_{\eta_0}^\eta \left[
  \hat{H}^{(2)}_\textsc{i}(\tau)-\tilde{H}^{(2)}_{\alpha_\textsc{b}(\tau)}
  \right] \dd \tau,
\label{perturV2}\\
\hat{C}'_2 &= \int_{\eta_0}^\eta \dd \tau_1 \int_{\eta_0}^\eta \dd
\tau_2 \hat{H}^{(2)}_\textsc{i}(\tau_1)
\tilde{H}^{(2)}_{\alpha_\textsc{b}(\tau_2)},
\label{perturV3}\\
 \hat{C}''_2 &= - \frac{1}{2} \int_{\eta_0}^\eta \dd \tau_1
 \int_{\eta_0}^\eta \dd \tau_2 \, T \left[
   \hat{H}^{(2)}_\textsc{i}(\tau_1)
   \hat{H}^{(2)}_\textsc{i}(\tau_2) +
   \tilde{H}^{(2)}_{\alpha_\textsc{b}(\tau_1)}
   \tilde{H}^{(2)}_{\alpha_\textsc{b}(\tau_2)} \right].
\label{perturV1}
\end{align}
\label{perturV}
\end{subequations}
The term $\hat{C}'_2$ comes from the mixed product of the two first
order terms of the time ordered operators, whereas the term
$\hat{C}''_2$ involves the sum of the two second order terms of the
time ordered operators. It turns out that the sum
$\hat{C}'_2+\hat{C}''_2$ can be put in a compact form as
\begin{equation}
\hat{C}_2 = \hat{C}'_2+\hat{C}''_2 = - \frac{1}{2} \int_{\eta_0}^\eta
\dd \tau_1 \int_{\eta_0}^\eta \dd \tau_2 \, T \left\{ \left[
  \hat{H}^{(2)}_\textsc{i}(\tau_1) -
  \tilde{H}^{(2)}_{\alpha_\textsc{b}(\tau_1)} \right] \left[
  \hat{H}^{(2)}_\textsc{i}(\tau_2) -
  \tilde{H}^{(2)}_{\alpha_\textsc{b}(\tau_2)} \right] \right\}.
\label{secondorder}
\end{equation}

\end{widetext}
Following the arguments developed in the previous sections, we define
the Beyond Born-Oppenheimer (BBO) perturbation state
$\ket{\psi_\textsc{p}^\textsc{(bbo)}(\eta)}$ as a projection of the
multiverse state (up to a normalization factor):
\begin{align}
\ket{\psi_\textsc{p}^\textsc{(bbo)}(\eta)} \eqdef
\hat{\Pi}_{\alpha_\textsc{b}(\eta)} \ket{\Phi(\eta)} =
\hat{\Pi}_{\alpha_\textsc{b}(\eta)} \hat{V}(\eta, \eta_0)
\ket{\Phi^{(\textsc{bo})}(\eta)},
\end{align}
where $\alpha_\textsc{b}(\eta)$ is always the background state
trajectory evolving from $\alpha_\textsc{b}(\eta_0)$ with the
zeroth-order Hamiltonian $\hat{H}^{(0)}$. Making use of
$\hat{\Pi}_{\alpha_\textsc{b}(\eta)}$ and its adjoint, we obtain
\begin{equation}
\ket{\psi_\textsc{p}^\textsc{(bbo)}(\eta)} =
\hat{\Pi}_{\alpha_\textsc{b}(\eta)} \hat{V}(\eta, \eta_0)
\hat{\Pi}_{\alpha_\textsc{b}(\eta)}^\dagger
\,\ket{\psi_\textsc{p}^{(\textsc{bo})}(\eta)},
\end{equation}
which means that the difference between the two perturbation states
$\ket{\psi_\textsc{p}^{(\textsc{bo})}(\eta)}$ and
$\ket{\psi_\textsc{p}^\textsc{(bbo)}(\eta)}$ lies in the partial
expectation value of $\hat{V}(\eta,\eta_0)$ on the physical
background, i.e.,
\begin{equation}
\hat{\Pi}_{\alpha_\textsc{b}(\eta)} \hat{V}(\eta, \eta_0)
\hat{\Pi}_{\alpha_\textsc{b}(\eta)}^\dagger \equiv
\bra{\alpha_\textsc{b}(\eta)} \hat{V}(\eta,\eta_0)
\ket{\alpha_\textsc{b}(\eta)},
\label{AvgV2}
\end{equation}
the remark discussed below Eq.~\eqref{AvgH2} applying to
\eqref{AvgV2}.  In this framework, the actual background of our
Universe $\ket{\alpha_\textsc{b}(\eta)}$ plays a role similar to
vacuum in quantum field theory and the expectation value of
$\hat{V}(\eta,\eta_0)$ represents the dynamical effect of virtual
backgrounds on the Born-Oppenheimer perturbation state. It is very
similar to loop corrections, and in fact the perturbative calculations
developed in the following involve loop-calculations on virtual
backgrounds reminiscent of Feynman diagrams.

Finally, using the second order perturbative expression of
$\hat{V}(\eta,\eta_0)$ obtained above, we obtain
\begin{align}
\ket{\psi_\textsc{p}^\textsc{(bbo)}(\eta)} \simeq \left[\mathbbm{1} +
  \hat{\Pi}_{\alpha_\textsc{b}(0)} \left(\hat{C}_1+\hat{C}_2 \right)
  \hat{\Pi}_{\alpha_\textsc{b}(0)}^\dagger \right]
\ket{\psi_\textsc{p}^{(\textsc{bo})}(\eta)},
\label{BBOstate}
\end{align}
where we use that $\hat{\Pi}_{\alpha_\textsc{b}}
\hat{\Pi}_{\alpha_\textsc{b}}^\dagger =
\mathbbm{1}_{\mathcal{H}_\textsc{p}}$, $\ket{\alpha_\textsc{b}(\eta)}
= e^{-i \hat{H}^{(0)} \eta} \ket{\alpha_\textsc{b}(0)}$ and
$\hat{\Pi}_{\alpha_\textsc{b}(\eta)} =
\hat{\Pi}_{\alpha_\textsc{b}(0)} e^{i \hat{H}^{(0)} \eta}$.

\subsection{First-order correction}

From the definition of $\hat{C}_1$ in Eq. \eqref{perturV1}, we get its
contribution to Eq.~\eqref{BBOstate} by evaluating
$\hat{\Pi}_{\alpha_\textsc{b}(0)} \hat{C}_1
\hat{\Pi}_{\alpha_\textsc{b}(0)}^\dagger$.  Taking into account the
fact that, since $\tilde{H}^{(2)} _{\alpha_\textsc{b}(\tau)}$ acts
only on the perturbation Hilbert space, one has
\begin{equation}
\hat{\Pi}_{\alpha_\textsc{b}(0)}
\tilde{H}^{(2)}_{\alpha_\textsc{b}(\tau)}
\hat{\Pi}_{\alpha_\textsc{b}(0)}^\dagger
=\tilde{H}^{(2)}_{\alpha_\textsc{b}(\tau)}
\end{equation}
as well as
\begin{align}
\hat{\Pi}_{\alpha_\textsc{b}(0)} \hat{H}^{(2)}_\textsc{i}(\tau)
\hat{\Pi}_{\alpha_\textsc{b}(0)}^\dagger =
\hat{\Pi}_{\alpha_\textsc{b}(\tau)} \hat{H}^{(2)}
\hat{\Pi}_{\alpha_\textsc{b}(\tau)}^\dagger
=\tilde{H}^{(2)}_{\alpha_\textsc{b}(\tau)},
\end{align}
we find that the two contributions are equal and opposite, so that the
first-order correction in $\hat{H}^{(2)}$ from the virtual backgrounds
to the Born-Oppenheimer state is identically vanishing, independently
of the actual form of the background potential $\hat{V}_\textsc{b}$
coupling the perturbation and the background.

\subsection{Second-order correction (one-loop correction)}
\label{secondordersection}

We now calculate the effect of the term $\hat{C}_2$ defined in
Eq.~\eqref{secondorder} by using the explicit expression for
$\hat{H}^{(2)}=\sum_{\bm{k}} \hat{H}^{(2)}_{\bm{k}}$.
Eq.~\eqref{globalclassH} then yields
\begin{equation}
\tilde{H}^{(2)}_{\alpha_\textsc{b}(\eta)} = \sum_{\bm{k}}
\left\{
\hat{\pi}_{\bm{k}} \hat{\pi}_{-\bm{k}} + \left[
k^2 - V_\textsc{b}(\eta) \right] 
\hat{v}_{\bm{k}} \hat{v}_{-\bm{k}}
\right\}
\label{H2alpha}
\end{equation}
and
\begin{equation}
\hat{H}^{(2)}_\textsc{i}(\eta) = \sum_{\bm{k}} \left\{
\hat{\pi}_{\bm{k}} \hat{\pi}_{-\bm{k}} + \left[ k^2 -
  \hat{V}_\textsc{b}^{(\textsc{i})}(\eta)\right] \hat{v}_{\bm{k}}
  \hat{v}_{-\bm{k}}
  \right\},
\label{H2I}
\end{equation}
where
\begin{equation}
V_\textsc{b}(\eta) =\bra{\alpha_\textsc{b}(\eta)}
\hat{V}_\textsc{b}
\ket{\alpha_\textsc{b}(\eta)}
\label{VBfunc}
\end{equation}
is a function of time, the operator $\hat{V}_\textsc{b}$ depending
only on background variables, and
\begin{equation}
\hat{V}_\textsc{b}^{(\textsc{i})}(\eta) = e^{i \hat{H}^{(0)} \eta}
\hat{V}_\textsc{b} e^{-i \hat{H}^{(0)} \eta}.
\label{VBinteract}
\end{equation}
The only difference between \eqref{H2alpha} and \eqref{H2I} coming
from the potentials \eqref{VBfunc} and \eqref{VBinteract}, one easily
obtains the second-order term $\hat{C}_2$ as
\begin{widetext}
\begin{equation}
\hat{C}_2 = -\frac{1}{2} \left( \sum_{\bm{k}} \hat{v}_{\bm{k}}
\hat{v}_{-\bm{k}} \right)^2
\int_{\eta_0}^\eta \dd \tau_1
\int_{\eta_0}^\eta \dd \tau_2 \, T \left\{ \left[
  \hat{V}_\textsc{b}^{(\textsc{i})} (\tau_1) - V_\textsc{b} (\tau_1)
  \right]\left[ \hat{V}_\textsc{b}^{(\textsc{i})} (\tau_2) -
  V_\textsc{b} (\tau_2) \right] \right\},
\end{equation}
which, in view of the fact that $V_\textsc{b}(\eta)$ is a mere
function and thus commutes with all operators, can be transformed into
the simplified formula
\begin{equation}
\hat{C}_2 = -\frac{1}{2} \left( \sum_{\bm{k}} \hat{v}_{\bm{k}}
\hat{v}_{-\bm{k}} \right)^2 \int_{\eta_0}^\eta \dd \tau_1
\int_{\eta_0}^\eta \dd \tau_2 \, \left\{ T \left[
  \hat{V}_\textsc{b}^{(\textsc{i})} (\tau_1)
  \hat{V}_\textsc{b}^{(\textsc{i})} (\tau_2)\right] - V_\textsc{b}
(\tau_1) V_\textsc{b} (\tau_2) \right\}.
\end{equation}
Upon expanding the time ordering operator, this yields
\begin{equation}
\hat{C}_2 = -\left( \sum_{\bm{k}} \hat{v}_{\bm{k}}
\hat{v}_{-\bm{k}} \right)^2
\left\{ \left[ \int_{\eta_0}^\eta \dd
  \tau_1 \int_{\eta_0}^{\tau_1} \dd \tau_2 \,
  \hat{V}_\textsc{b}^{(\textsc{i})} (\tau_1)
  \hat{V}_\textsc{b}^{(\textsc{i})} (\tau_2) \right] - \frac{1}{2}
\left[ \int_{\eta_0}^\eta V_\textsc{b} (\tau) \dd\tau \right]^2
\right\}.
\end{equation}
Defining $K_{\eta,\eta_0} \in \mathbb{C}$ through
\begin{align}
 \left( \sum_{\bm{k}} \hat{v}_{\bm{k}}
\hat{v}_{-\bm{k}} \right)^2
 K_{\eta,\eta_0} = -\hat{\Pi}_{\alpha_\textsc{b}(0)}
 \hat{C}_2 \, \hat{\Pi}_{\alpha_\textsc{b}(0)}^\dagger,
\end{align}
we obtain
\begin{align}
K_{\eta,\eta_0}= \int_{\eta_0}^\eta \dd \tau_1 \int_{\eta_0}^{\tau_1}
\dd \tau_2 \, \bra{\alpha_\textsc{b}(0)}
\hat{V}_\textsc{b}^{(\textsc{i})} (\tau_1)
\hat{V}_\textsc{b}^{(\textsc{i})} (\tau_2) \ket{\alpha_\textsc{b}(0)}
- \frac{1}{2} \left[ \int_{\eta_0}^\eta V_\textsc{b} (\tau_1) d\tau_1
  \right]^2,
\label{Kintegral}
\end{align}
\end{widetext}
with which we finally get the (not normalized) BBO perturbation state
of \eqref{BBOstate} as
\begin{equation}
\ket{\psi_\textsc{p}^\textsc{(bbo)}(\eta)} \simeq \left[ \mathbbm{1} -
  K_{\eta,\eta_0} \left( \sum_{\bm{k}} \hat{v}_{\bm{k}}
  \hat{v}_{-\bm{k}} \right)^2 \right]
\ket{\psi_\textsc{p}^{(\textsc{bo})}(\eta)}.
\label{BBOfinal1}
\end{equation}
As the perturbative framework we work in is valid only if the
correction to the BO state is small, we must verify that
$\scalar{\psi_\textsc{p}^\textsc{(bbo)}(\eta)}{\psi_\textsc{p}
  ^\textsc{(bbo)}(\eta)} \simeq 1$. Equivalently, the condition
\begin{align}\label{valid}
|\Rez \, K_{\eta,\eta_0} | \left\langle\left( \sum_{\bm{k}} \hat{v}_{\bm{k}}
\hat{v}_{-\bm{k}}
\right)^2\right\rangle^{(\textsc{bo})}_\eta \ll 1,
\end{align}
where $~\rangle^{(\textsc{bo})}_\eta:=\ket{\psi_\textsc{p}
  ^{(\textsc{bo})}(\eta)}$, must be fulfilled. From Wick's (or,
Isserlis') theorem, in the case of zero-mean Gaussian distribution,
which we expect to be valid in the BO approximation, we obtain
\begin{align}\label{integrals}\begin{split}
\left\langle\left( \sum_{\bm{k}} \hat{v}_{\bm{k}} \hat{v}_{-\bm{k}}
\right)^2\right\rangle^{(\textsc{bo})}_\eta = &
\left[\left\langle\sum_{\bm{k}} \hat{v}_{\bm{k}} \hat{v}_{-\bm{k}}
  \right\rangle^{(\textsc{bo})}_\eta\right]^2\\ & +
2\sum_{\bm{k}}\left[\left\langle \hat{v}_{\bm{k}} \hat{v}_{-\bm{k}}
  \right\rangle^{(\textsc{bo})}_\eta\right]^2.
\end{split}
\end{align}
The power spectrum $P_\eta(k)$ is usually defined by the 2$-$point
function of $\zeta$, the so-called curvature perturbation, which is
related to $v$ through a background time-dependent factor. Defining
the discrete Dirac delta as $\delta_{\bm{v}\not= \bm{0}} = 0$ and
$\delta_{\bm{0}} = 1$, we have $\langle \hat{v}_{\bm{k}}
\hat{v}_{\bm{\ell}}\rangle_\eta = \delta_{\bm{k}+\bm{\ell}} f(\eta)
P_\eta(k)$; the function of time $f(\eta)$ is unknown at this stage as
it depends on the actual model (inflation, bounce,...) under
consideration. For the purpose of setting the above formalism, its
precise form is not necessary here. The primordial power spectrum is
obtained at some later stage at which $P_\eta(k)$ becomes
time-independent and satisfies $P_\eta(k)\to P(k)\propto
k^{n_\textsc{s}-4}$. To actually perform the calculations needed in
e.g., Eq.~\eqref{integrals}, one can take the continuous limit
obtained by replacing the discrete $\delta_{\bm{v}+\bm{w}}$ by the
actual Dirac distribution, i.e.
$$
\delta_{\bm{v}+\bm{w}} \to (2\pi)^3 \delta^{(3)}(\bm{v}+ \bm{w}),
$$
while the discrete sum transforms in the corresponding integral,
namely
$$\sum_{\bm{v}} \to \frac{1}{(2\pi)^3} \int \dd^3 \bm{v}.$$
Both the sums and the corresponding integrals usually diverge in
ordinary Minkowski space, and need be regularized. In the cosmological
setup under discussion here, one can appeal to the finite size of the
universe to restrict the bounds in $k-$space, thereby defining
\begin{equation}
S_\textsc{r} = \left\langle\sum_{\bm{p}} \hat{v}_{\bm{p}} \hat{v}
_{-\bm{p}}\right\rangle^{(\textsc{bo})}_\eta,
\label{SR}
\end{equation}
a regularized quantity. Note that in many cases of cosmological
interest, this quantity contains a positive power of the scale factor,
which can be large. We subsequently assume that $S_\textsc{r}\gg
\langle \hat{v}_{\bm{k}} \hat{v}_{-\bm{k}} \rangle$ (for a given fixed
$\bm{k}$) in what follows, which allows us to keep only the leading
terms.

\section{Primordial signatures of the multiverse  dynamics}
\label{PNGsec}

Let us now compute the (equal-time) correlation functions for the BBO
state in order to determine the effect of the multiverse dynamics on
the primordial structure. Within the BO approach, the perturbation's
dynamics starts from a vacuum state and ends in a final Gaussian
state. For the latter only even-point correlation functions are
non-vanishing and are fully expressible in terms of products of
$2-$point functions.  We find it sufficient to consider the
leading-order $2-$point, $3-$point and $4-$point functions.

\subsection{Power spectrum}

The power spectrum of the perturbations in the BBO state is given by
the $2-$point function
\begin{equation}
\langle \hat{v}_{\bm{k}}\hat{v}_{\bm{\ell}}
\rangle^\textsc{(bbo)}_\eta =
\frac{\bra{\psi_\textsc{p}^\textsc{(bbo)}(\eta)}
  \hat{v}_{\bm{k}}\hat{v}_{\bm{\ell}}
  \ket{\psi_\textsc{p}^\textsc{(bbo)}(\eta)}}{\scalar{\psi_\textsc{p}^\textsc{(bbo)}
    (\eta)}{\psi_\textsc{p}^\textsc{(bbo)}(\eta)}},
\end{equation}
for two wave vectors $\bm{k}$ and $\bm{\ell}$. Using the BBO expansion
\eqref{BBOfinal1}, one gets
\begin{widetext}
\begin{equation}
\bra{\psi_\textsc{p}^\textsc{(bbo)}(\eta)}
\hat{v}_{\bm{k}}\hat{v}_{\bm{\ell}}
\ket{\psi_\textsc{p}^\textsc{(bbo)}(\eta)} \simeq
\left\langle\hat{v}_{\bm{k}}\hat{v}_{\bm{\ell}}\right\rangle^{(\textsc{bo})}_\eta-
2\, \Rez \left(K_{\eta, \eta_0} \right)
\left\langle\hat{v}_{\bm{k}}\hat{v}_{\bm{\ell}}\left( \sum_{\bm{p}}
\hat{v}_{\bm{p}} \hat{v}_{-\bm{p}}
\right)^2\right\rangle^{(\textsc{bo})}_\eta,
\end{equation}
as well as
\begin{equation}
\scalar{\psi_\textsc{p}^\textsc{(bbo)}(\eta)}{\psi_\textsc{p}
  ^\textsc{(bbo)}(\eta)} \simeq 1 - 2\, \Rez \left(K_{\eta, \eta_0}
\right) \left\langle\left( \sum_{\bm{p}} \hat{v}_{\bm{p}}
\hat{v}_{-\bm{p}} \right)^2\right\rangle^{(\textsc{bo})}_\eta,
\end{equation}
leading finally to
\begin{equation}
\langle \hat{v}_{\bm{k}}\hat{v}_{\bm{\ell}}
\rangle^\textsc{(bbo)}_\eta \simeq
\langle\hat{v}_{\bm{k}}\hat{v}_{\bm{\ell}}\rangle^{(\textsc{bo})}_\eta-
2\, \Rez \left(K_{\eta, \eta_0} \right)\left[
  \left\langle\hat{v}_{\bm{k}}\hat{v}_{\bm{\ell}}\left( \sum_{\bm{p}}
  \hat{v}_{\bm{p}} \hat{v}_{-\bm{p}}
  \right)^2\right\rangle^{(\textsc{bo})}_\eta-\langle\hat{v}_{\bm{k}}\hat{v}_{\bm{\ell}}
  \rangle^{(\textsc{bo})}_\eta \left\langle\left( \sum_{\bm{p}}
  \hat{v}_{\bm{p}} \hat{v}_{-\bm{p}}
  \right)^2\right\rangle^{(\textsc{bo})}_\eta\right],
\label{Pk}
\end{equation}
from which one can derive the corrections due to the virtual states.

Eq.~\eqref{Pk} is given as a sum of moments of the (zero mean)
multivariate normal distribution. As done above, by virtue of Wick's
theorem, it can be expressed in terms of products of all possible
contractions into second-order expectations values. In particular, the
square bracket becomes
\begin{equation}
  4\sum_{\bm{p},\bm{q}}\left[ \langle\hat{v}_{\bm{k}}\hat{v}_{\bm{q}}
    \rangle^{(\textsc{bo})}_\eta \langle
    \hat{v}_{\bm{\ell}}\hat{v}_{-\bm{q}}\rangle^{(\textsc{bo})}_\eta
    \langle \hat{v}_{\bm{p}} \hat{v}_{-\bm{p}}
    \rangle^{(\textsc{bo})}_\eta +2\langle
    \hat{v}_{\bm{k}}\hat{v}_{\bm{p}} \rangle^{(\textsc{bo})}_\eta
    \langle \hat{v}_{\bm{\ell}}\hat{v}_{-\bm{q}}
    \rangle^{(\textsc{bo})}_\eta \langle \hat{v}_{-\bm{p}}
    \hat{v}_{\bm{q}} \rangle^{(\textsc{bo})}_\eta \right],
\end{equation}
\end{widetext}
in which the second term can be omitted as it is expected to be
negligible with respect to the first one proportional to
$S_\textsc{r}$ according to our hypothesis below \eqref{SR}.  After
defining the two power spectra $P^\textsc{(i)}_\eta(k)$ (with
\textsmaller{I}$=$\textsmaller{BO} or \textsmaller{BBO}), the final
result thus reads
\begin{equation}
P^\textsc{(bbo)}_\eta(k) = P^\textsc{(bo)}_\eta(k) \left[ 1 - 8\Rez
  \left(K_{\eta, \eta_0} \right) S_\textsc{r} P^\textsc{(bo)}_\eta(k)
  \right],
\label{PS}
\end{equation}
showing a nonlinear modification of the power spectrum; this gives a
serious hint that non-Gaussian features should consequently appear, to
which we now turn.

\subsection{Bispectrum}

The $3-$point function, defined through
\begin{equation}
\langle \hat{v}_{\bm{k}}\hat{v}_{\bm{\ell}}\hat{v}_{\bm{m}}
\rangle^\textsc{(bbo)}_\eta =
\frac{\bra{\psi_\textsc{p}^\textsc{(bbo)}(\eta)}
  \hat{v}_{\bm{k}}\hat{v}_{\bm{\ell}}\hat{v}_{\bm{m}}
  \ket{\psi_\textsc{p}^\textsc{(bbo)}(\eta)}}
     {\scalar{\psi_\textsc{p}^\textsc{(bbo)}
         (\eta)}{\psi_\textsc{p}^\textsc{(bbo)}(\eta)}},
\end{equation}
yields the so-called bispectrum~\cite{Lewis:2011au}. Upon expanding as
for the spectrum, one finds that all the relevant terms contain odd
numbers of $\hat{v}_{\bm{k}}$ operators. Using Wick's theorem again,
one gets that each terms contains an expectation value alone, which is
vanishing by construction.  As a result, the bispectrum is found to
vanish in the BBO state.

Note that this conclusion holds true for all odd$-$point functions.

\subsection{Trispectrum}

\begin{figure}[t]
\centering
\includegraphics[width=0.4\textwidth]{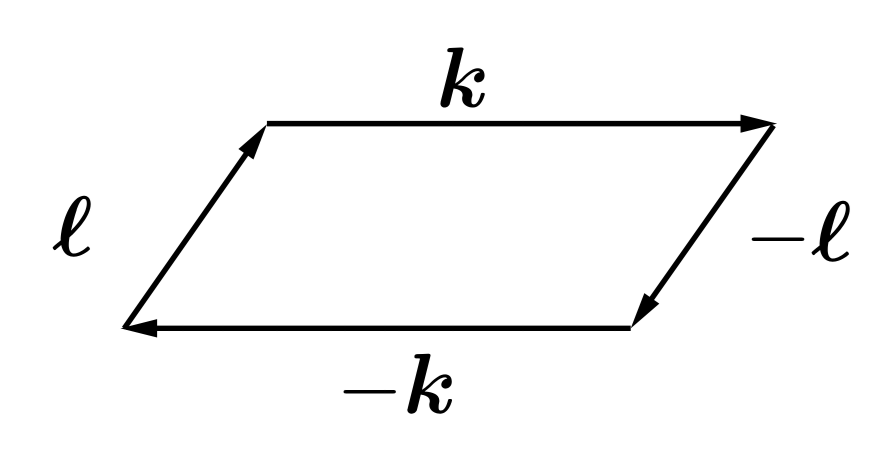}

\caption{Schematic illustration of the parallelogram shape of the
  trispectrum in the Fourier space.}

\label{fig:trisp}
\end{figure}

Going one step further, we can compute the $4-$point function (or any
higher-order even$-$point function) for the BBO state in a fashion
similar to the power spectrum, namely writing
\begin{equation}
\langle
\hat{v}_{\bm{k}}\hat{v}_{\bm{\ell}}\hat{v}_{\bm{m}}\hat{v}_{\bm{n}}
\rangle^\textsc{(bbo)}_\eta =
\frac{\bra{\psi_\textsc{p}^\textsc{(bbo)}(\eta)}
  \hat{v}_{\bm{k}}\hat{v}_{\bm{\ell}}\hat{v}_{\bm{m}}\hat{v}_{\bm{n}}
  \ket{\psi_\textsc{p}^\textsc{(bbo)}(\eta)}}
     {\scalar{\psi_\textsc{p}^\textsc{(bbo)}
         (\eta)}{\psi_\textsc{p}^\textsc{(bbo)}(\eta)}}.
\end{equation}
Using again Eq.~\eqref{BBOfinal1}, this relation can be expanded as
before in terms of the BO states. Doing so, one gets, to leading
order,
\begin{equation}
\langle
\hat{v}_{\bm{k}}\hat{v}_{\bm{\ell}}\hat{v}_{\bm{m}}\hat{v}_{\bm{n}}
\rangle^\textsc{(bbo)}_\eta \simeq
\langle\hat{v}_{\bm{k}}\hat{v}_{\bm{\ell}}\hat{v}_{\bm{m}}\hat{v}_{\bm{n}}
\rangle^{(\textsc{bo})}_\eta+T({\bm{k}},{\bm{\ell}},{\bm{m}},{\bm{n}}),
\end{equation}
where $T({\bm{k}},{\bm{\ell}},{\bm{m}},{\bm{n}})$ is the so-called
trispectrum, which is the connected part of the $4-$point function.
Note that $\langle\hat{v}_{\bm{k}}\hat{v}_{\bm{\ell}}\hat{v}
_{\bm{m}}\hat{v}_{\bm{n}}\rangle^{(\textsc{bo})}_\eta$ is the purely
unconnected part so that, as any even$-$point function, it is fully
expressible in terms of products of $2-$point functions in the
Gaussian state. The trispectrum $T({\bm{k}},{\bm{\ell}},{\bm{m}},
{\bm{n}})$ is therefore related with actual non-Gaussianities which we
are aiming at. The same leading-order approximation as applied to get
the spectrum Eq.~\eqref{Pk} now provides
\begin{widetext}
\begin{equation}
T({\bm{k}},{\bm{\ell}},{\bm{m}},{\bm{n}})= -2\, \Rez \left(K_{\eta,
  \eta_0} \right) \left[ \left\langle
  \hat{v}_{\bm{k}}\hat{v}_{\bm{\ell}}\hat{v}_{\bm{m}}
  \hat{v}_{\bm{n}}\left( \sum_{\bm{p}} \hat{v}_{\bm{p}} \hat{v}
  _{-\bm{p}} \right)^2\right\rangle^{(\textsc{bo})}_\eta -
  \langle\hat{v}_{\bm{k}} \hat{v}_{\bm{\ell}} \hat{v}_{\bm{m}}
  \hat{v}_{\bm{n}} \rangle^{(\textsc{bo})}_\eta \left\langle \left(
  \sum_{\bm{p}} \hat{v}_{\bm{p}} \hat{v} _{-\bm{p}} \right)^2
  \right\rangle^{(\textsc{bo})}_\eta\right],
\label{TS2}
\end{equation}
which, upon applying Wick's theorem, yields 105 terms for the first
expectation value, among which the 9 of the second part, leading to a
total of 96 terms. It takes the form of $T\sim T_9 S_\textsc{r}^2 +
T_{24} S_\textsc{r} + T_{72}$, the index representing the number of
combinations in each factor; the cancellation concerns the terms
proportional to $S_\textsc{r}^2$, which vanishes. After retaining only
those with the factor $S_\textsc{r}$ and simplifying the resulting
expressions using that $\langle\hat{v}_{\bm{k}}\hat{v}
_{\bm{\ell}}\rangle^{(\textsc{bo})}_\eta \propto \delta_{\bm{k}+
  \bm{\ell}}$, we obtain
\begin{align}\label{TS}
\begin{split}
T({\bm{k}},{\bm{\ell}},{\bm{m}},{\bm{n}})=&-4\, \Rez \left(K_{\eta,
  \eta_0} \right)S_\textsc{r} \delta_{\bm{k}+\bm{\ell}+\bm{m}+\bm{n}}
\left\{ \langle\hat{v}_{\bm{k}} \hat{v}_{-\bm{k}}
\rangle^{(\textsc{bo})}_\eta \left[ \langle \hat{v}_{\bm{m}}
  \hat{v}_{\bm{-m}} \rangle^{(\textsc{bo})}_\eta \right]^2
(\delta_{\bm{k}+\bm{\ell}} +
\delta_{\bm{k}+\bm{n}})\right.\\ &\left. + \langle\hat{v}_{\bm{\ell}}
\hat{v}_{-\bm{\ell}} \rangle^{(\textsc{bo})}_\eta \left[
  \langle\hat{v}_{\bm{k}} \hat{v}_{-\bm{k}}
  \rangle^{(\textsc{bo})}_\eta \right]^2 (\delta_{\bm{\ell}+\bm{m}} +
\delta_{\bm{\ell}+\bm{n}}) + \langle \hat{v}_{\bm{n}}
\hat{v}_{\bm{-n}} \rangle^{(\textsc{bo})}_\eta \left[
  \langle\hat{v}_{\bm{\ell}} \hat{v}_{-\bm{\ell}}
  \rangle^{(\textsc{bo})}_\eta \right]^2 (\delta_{\bm{n}+\bm{m}} +
\delta_{\bm{n}+\bm{k}}) \right\}.
\end{split}
\end{align}
Given the form of the power spectrum \eqref{PS} and the form of the
trispectrum \eqref{TS}, we define the non-linear amplitude
$g^\textsc{bbo}_\textsc{nl} = -4 \Rez \left( K_{\eta, \eta_0} \right)
S_\textsc{r}$ and the shape of the trispectrum
$t({\bm{k}},{\bm{\ell}},{\bm{m}},{\bm{n}}):=T({\bm{k}},{\bm{\ell}},
{\bm{m}},{\bm{n}})/g^\textsc{bbo}_\textsc{nl}$.  The most general
shape is given by the parallelogram in the momentum space with two
pairs of conjugate momenta, and an arbitrary angle between these pairs
(see Fig. \ref{fig:trisp}). We note that it corresponds to the
parallelogram limit of the so-called ``local" shape that was
constrained by the Planck collaboration in
Ref.~\cite{Planck:2019kim}. For instance, in the limit
$\bm{\ell}\rightarrow -\bm{k}$ (implying $\bm{n}\rightarrow -
\bm{m}$), we obtain
\begin{equation}
t({\bm{k}},{-\bm{k}},{\bm{m}},{-\bm{m}}) =
\delta_{\bm{k}+\bm{\ell}+\bm{m}+\bm{n}}
\delta_{\bm{k}+\bm{\ell}}\left\{ \langle
\hat{v}_{\bm{k}}\hat{v}_{-\bm{k} } \rangle^{(\textsc{bo})}_\eta \left[
  \langle \hat{v}_{\bm{m}} \hat{v}_{\bm{-m}}
  \rangle^{(\textsc{bo})}_\eta \right]^2 + \langle
\hat{v}_{\bm{m}}\hat{v}_{\bm{-m}} \rangle^{(\textsc{bo})}_\eta \left[
  \langle
  \hat{v}_{\bm{k}}\hat{v}_{-\bm{k}}\rangle^{(\textsc{bo})}_\eta
  \right]^2 \right\},
\end{equation}
which is nothing but
$\delta_{\bm{k}+\bm{\ell}}t^\text{local}({\bm{k}},{\bm{\ell}},{\bm{m}},{\bm{n}})$,
where
$$
t^\text{local}({\bm{k}},{\bm{\ell}},{\bm{m}},{\bm{n}}) =
\delta_{\bm{k}+\bm{\ell}+\bm{m}+\bm{n}} \left[ \langle
  \hat{v}_{\bm{k}}\hat{v}_{-\bm{k}} \rangle^{(\textsc{bo})}_\eta
  \langle \hat{v}_{\bm{\ell}}
  \hat{v}_{-\bm{\ell}}\rangle^{(\textsc{bo})}_\eta \langle
  \hat{v}_{\bm{m}} \hat{v}_{\bm{-m}} \rangle^{(\textsc{bo})}_\eta +
  \text{3 perms} \right]
$$
is the ``local" shape given by Eq. (43) in
Ref.~\cite{Planck:2019kim}. Assuming the standard power spectrum $P(k)
\propto k^{n_s-4}$ then yields the {\sl universal} shape function
\begin{equation}
t({\bm{k}},{-\bm{k}},{\bm{\ell}},{-\bm{\ell}}) \propto \frac{1}{k^{4-
    n_s}} \left( \frac{1}{\ell^{4-n_s}} \right)^2 + \frac{1}{\ell^{4-
    n_s}} \left( \frac{1}{k^{4-n_s}} \right)^2 \propto [P(k)]^3 \left[
  \left( \frac{k}{\ell} \right)^{2(4-n_s)} + \left(\frac{k} {\ell}
  \right)^{4-n_s} \right]
\label{shape}
\end{equation}
\end{widetext}
depending essentially on the ratio $k/\ell$ between the lengths of the
parallelogram's sides in Fig.~\ref{fig:trisp}. The amplitude of the
trispectrum does not depend on the angle between the sides but only on
their relative lengths, and it scales with the overall size of the
parallelogram as the power spectrum cubed.

The Planck collaboration did not find evidence for a non-zero
primordial trispectrum and constrained the amplitude
$|g_\textsc{nl}^\text{local}|\lesssim 10^4-10^5$, which is therefore
for the moment largely unconstrained. However, the minimal prediction
we foresee in our context of quantizing the background as well as the
perturbations is that of a non vanishing signal, with an amplitude to
be determined in a model-dependent way, for the very specific
shape~\eqref{shape}.

Methods for analysing the shapes of non-Gaussianity with cosmic
microwave background and large scale structure data are
well-developed, in particular for the case of bispectrum (see,
e.g. \cite{Liguori:2010hx} for a review). The case of the CMB
trispectrum has been investigated in
\cite{Smith:2015uia,Hindmarsh:2009es} (see also
Ref.~\cite{AnilKumar:2022flx} for detection issues). On the other
hand, obtaining the theoretical value of the amplitude
$g^\textsc{bbo}_\textsc{nl}$ requires choosing a specific model and
calculating the value of $K(\eta,\eta_0)$ via
Eq. \eqref{Kintegral}. We stress again that this calculation can be
performed both for inflationary as well as alternative models,
potentially leading to distinct predictions.

\subsection{Heisenberg picture}

Some calculations may be more easily performed in the Heisenberg
picture. Making use of the definition
$\ket{\psi_\textsc{p}^\textsc{(bbo)}(\eta)} =
\hat{\Pi}_{\alpha_\textsc{b}(\eta)} \hat{V}(\eta, \eta_0)
\ket{\Phi^{(\textsc{bo})}(\eta)}$, we obtain the Heisenberg form of
$\hat{v}_k$, which we denote by $\hat{v}_k^\textsc{(bbo)}$, namely

\begin{equation}
\hat{v}_k^\textsc{(bbo)}= \hat{V}(\eta, \eta_0)^{\dagger}
\hat{\Pi}_{\alpha_\textsc{b}(\eta)}^{\dagger} \hat{v}_k
\hat{\Pi}_{\alpha_\textsc{b}(\eta)} \hat{V}(\eta, \eta_0).
\end{equation}
Upon using the expansion discussed above, we get
\begin{equation}
\hat{v}_k^\textsc{(bbo)}=\left[\hat{v}_k - 2\Rez \left(K_{\eta,
    \eta_0} \right) \left(\sum_{\bm{p}} \hat{v}_{\bm{p}} \hat{v}_{-
    \bm{p}}\right)^2\hat{v}_k\right],
\end{equation}
where we omitted the projector
$\hat{\Pi}_{\alpha_\textsc{b}(\eta)}^{\dagger}\hat{\Pi}_{\alpha_\textsc{b}(\eta)}$,
acting on $\mathcal{H}_\textsc{b}$. Switching to the position
representation, we finally get
\begin{align}
\hat{v}^\textsc{(bbo)}(x)=\hat{v}(x) - 2\Rez \left(K_{\eta, \eta_0}
\right) \left[\int\hat{v}^2(y)\dd^3y\right]^2\hat{v}(x),
\label{NGx}
\end{align}
where $\hat{v}(x)$ is a Gaussian field evolving according to the BO
dynamics. From Eq. \eqref{NGx}, it is evident that the non-Gaussianity
of $\hat{v}^\textsc{(bbo)}(x)$ is very non-local.

\section{Discussion}\label{Sumsec}

In this work we have derived a perturbation theory of the multiverse
dynamics of quantum cosmological systems, starting from the first
principles. We showed that the multiverse states develop inevitably in
the quantum models of primordial universe when the interaction between
the background and the perturbation is carefully taken into
account. The main result is the derivation of the general formulas
\eqref{BBOfinal1} and \eqref{Kintegral}, which give an analytical form
of the multiverse correction to the initial Born-Oppenheimer state of
the universe, that is, the tensor-product of a background state and an
adiabatic vacuum for the perturbation.

It was then shown that these new cosmological states describe the
primordial fluctuations with non-Gaussian features of a very specific
form. Namely, we found that the connected correlation functions of
even order do not vanish, and evaluated the the lowest-order
contribution, i.e. the trispectrum. We found that the its amplitude
depends on the scale-invariant function $K(\eta,\eta_0)$ and on the
power spectrum of the BO state. Assuming the latter to be
approximately scale-invariant, a universal shape function of the
trispectrum was derived. The numerical value of the expected amplitude
of non-Gaussianity, not calculable in a general way, must be obtained
by setting a specific model for the primordial universe. An earlier
evaluation was provided in a simplified bouncing
model~\cite{Bergeron:2024art} containing only two background
states. Whether similar conclusion will hold in generic inflationary
models still remains to be verified. Nevertheless, we expect that, had
a quantum gravity phase preceded an inflationary phase, the
non-Gaussianities produced by the quantum gravity phase should have
survived the inflation and may not have been simply washed out.

The density fluctuations inside other branches of the primeval
multiverse alter via gravitational interaction the geometry of our
Universe, causing small but in principle measurable curvature
imprints. When superimposed on the linear structure, these imprints
produce a distinct pattern of non-Gaussianities, making the multiverse
scenario a testable physical theory. It must be noted, however, that
no signal of non-Gaussianities has been detected so far. The Planck
collaboration has put constraints on some shapes of the bispectrum and
the trispectrum \cite{Planck:2019kim}. However, these constraints do
not concern the trispectrum obtained from the multiverse dynamics and
a new analysis of the available data involving the obtained shape
function is necessary. Various ongoing and planned CMB experiments
will significantly improve polarization sensitivity and measurements
down to smaller scales, further constraining non-Gaussianities
\cite{CMB-S4:2016ple, SimonsObservatory:2018koc, Alvarez:2019rhd}.

Let us note that a potential detection of the non-Gaussianities of the
predicted shape would strongly enforce the evidence for quantum
gravity effects in the primordial universe despite the fact that the
primordial gravity waves that are widely held as a clean quantum
gravity signature, remain undetected.

The next step in the development of the multiverse formalism is to
determine the explicit value of the function $K(\eta,\eta_0)$ defined
in Eq. \eqref{Kintegral}. We emphasize that the developed perturbation
theory does not assume any specific quantum cosmological model and
hence, the formula \eqref{Kintegral} can be evaluated in various
ways. In subsequent forthcoming work, we will apply our present
formalism to a bouncing model based on analytically determined quantum
background solutions~\cite{Bergeron:2023zzo} as well as to a simple
inflationary model.

Let us finish this work by noting that the existence of multiverse
states, though completely correct and even necessary from the point of
view of quantum formalism, raise difficult interpretational
issues. The existence of multiple branches in the primordial
multiverse, and their imprints on the observed cosmological
fluctuations in our branch, prompt questions such as: Did these other
branch universes really exist? Do they still exist? Is the wave
function a tangible, physical entity or, is it merely a mathematical
construct that should not be confused with the physical reality?
Presently, none of these questions can be really answered, however, it
is plausible that our framework could to some extent be useful in
addressing some of these and similar issues.

\begin{acknowledgements}

P.M. acknowledges the support of the National Science Centre (NCN,
Poland) under the Research Grant 2018/30/E/ST2/00370.
\end{acknowledgements}

\bibliography{refs.bib}

\end{document}